\documentclass[useAMS,usenatbib]{mn2e}
\usepackage{times}
\usepackage{natbib}
\usepackage{lscape}
\usepackage[usenames]{color}
\usepackage{graphicx}
\usepackage{amssymb}
\usepackage{pifont}
\usepackage{ulem}
\usepackage{amsmath}
\usepackage{fixltx2e}

\renewcommand\appendix{\par
  \setcounter{section}{0}
  \setcounter{subsection}{0}
  \setcounter{figure}{0}
  \setcounter{table}{0}
  \renewcommand\thesection{APPENDIX \Alph{section}}
  \renewcommand\thefigure{\Alph{section}\arabic{figure}}
  \renewcommand\thetable{\Alph{section}\arabic{table}}
}

\long\def\Ignore#1{\relax}

\newcommand{\degrees}{$^\circ$}

\newcommand{\ie}{{\it i.e.}}

\title[The CDFS as a test case for GMCAO]{The {\it Chandra} Deep Field South as a test case for Global-MCAO}
\author[E. Portaluri et al.]
          {E. Portaluri,$^{1,2}$\thanks{E-mail:elisa.portaluri@oapd.inaf.it}
    V. Viotto,$^{1,2}$
    R. Ragazzoni,$^{1,2}$
    M. Gullieuszik,$^{1,2}$
    M. Bergomi,$^{1,2}$\and
    D. Greggio,$^{1,2,3}$
    F. Biondi,$^{1,2}$
    M. Dima,$^{1,2}$
    D. Magrin,$^{1,2}$ and
    J. Farinato$^{1,2}$\\
\\$^{1}$INAF--Osservatorio Astronomico di Padova, vicolo dell'Osservatorio 5, I-35122 Padova, Italy
\\$^{2}$ADONI - Laboratorio Nazionale Ottiche Adattive - Italy
\\$^{3}$Dipartimento di Fisica e Astronomia `G. Galilei', Universit\`a degli Studi di Padova, vicolo dell'Osservatorio 3, I-35122 Padova, Italy}

\begin{document}

\date{Accepted xxx Received xxx ; in original form \today}

\maketitle

\label{firstpage}

\begin{abstract}
The era of the next generation of giant telescopes requires not only the advent of new technologies but also the development of novel methods in order to exploit fully the extraordinary potential they are built for.
Global Multi Conjugate Adaptive Optics (GMCAO) pursues this approach, with the goal of achieving good performance over a few arcmin field of view and an increase in sky coverage.
In this article we show the gain offered by this technique to an astrophysical application, such as the photometric survey strategy applied to the {\it Chandra} Deep Field South as a case study. We simulated a close-to-real observation of a 500 $\times$ 500 arcsec$^2$ extragalactic deep field with a 40-m class telescope that implements GMCAO.
We analysed mock $K$-band images of the 6000 high-redshift (up to $z$=2.75) galaxies therein as if they were real to recover the initial input parameters. We attained 94.5 per cent completeness for the source detection with SExtractor. We also measured the morphological parameters of all the sources with the 2-dimensional fitting tools GALFIT. The agreement we found between the recovered and the intrinsic parameters demonstrates GMCAO as a reliable approach to assist extremely large telescope (ELT) observations of extragalactic interest.
\end{abstract}

\section{Introduction}  
\label{sec:intro}
An extensive way to study the formation and evolution of galaxies and their association in groups or clusters is via deep multi-band observation campaigns, carried out either from space or with ground-based telescopes. They offer researchers the chance to investigate the large-scale structures in a selected region of the sky over a broad interval of look-back time.
In the last 20 years, this technique has represented the most promising way to probe the distant Universe, translating into a considerable advance in our knowledge of high-redshift objects, answering fundamental questions and opening new and often more substantial ones.
As new technologies and capabilities were reached, new instruments were designed to prove the ability of available telescopes to probe even deeper limits.
In fact, the combination of deeper images and new cameras with sensitivity at redder wavelengths has pushed the redshift limits well into the reionization epoch of the Universe.

The initial step towards this goal was the observation in 1995 of the {\it Hubble Deep Field} \citep{Williams1996}. Afterwards, this region was extensively studied by numerous programs and surveys, ranging from optical to near infrared domain. Then there was also the {\it Hubble Ultra Deep Field} \citep{Beckwith2006}, together with the wide-field Great Observatories Origins Deep Surveu (GOODS) images \citep{Giavalisco2004}, which revealed for the first time $z\approx$ 6 galaxies at the end of the reionization epoch \citep{Bouwens2006}; the ultra-deep coverage in the near-infrared of the {\it Hubble Ultra Deep Field 2009}, HUDF09 \citep{Bouwens2011}, which allowed for the first systematic exploration of galaxies at z $\approx$ 10; the wide-field multi-wavelength catalog of Cosmic Assembly Near-infrared Deep Extragalactic Legacy Survey, CANDELS \citep{Galametz2013}, which improved the statistics at higher luminosities, and the near-IR counterpart {\it Hubble Ultra Deep Field 2012}, HUDF12 \citep{Ellis2013}.
Moreover, deep X-ray observations of the {\it Chandra} Deep Field South (CDFS, \citealt{Giacconi2000}) proved the great interest of the community in unravelling the assembly history that shapes the objects observed today, and advances this region as one of the primary sites for cosmological analysis, motivating observations from the X-ray to radio wavelengths, both from space- and ground-based observatories.

In order to make more substantial progress in this field, and to improve the potential of future extremely large telescopes (ELTs), such as the the European Extremely Large Telescope \citep{Gilmozzi2007}, the Giant Magellan Telescope \citep{Johns2008}, and the Thirty Meter Telescope \citep{Szeto2008}, one can conceive and actualise novel concepts and new ideas. As an exciting challenge, \citet{Ragazzoni2010} proposed an extension of the Multi Conjugate Adaptive Optics technique (MCAO), called Global-MCAO (GMCAO), in which a wide field of view (labelled as technical FoV) can be used to look for natural guide stars while the correction is performed on a smaller, but adequate scientific FoV. It combines a number of concepts of Adaptive Optics (AO), some of them already well known or even proven on the sky, working in a numerical layer-oriented fashion, consisting of several deformable mirrors conjugated to different layers of the turbulence, instead of integrating over a single direction. 
A review of the implementation of an example of such a system can be found in \citet{Viotto2015}. The fact that GMCAO uses only natural guide stars translates into the importance of extending the area where the stars can be found, leading to an increase in sky coverage - one of the arguments, together with the uniformity of the corrected FoV, in favour of implementing a laser guide stars (LGSs)-based system \citep{Rigaut2014}. With GMCAO, the use of LGSs can be avoided, freeing this method from the associated issues \citep{Fried1995,Pfrommer2009,Diolaiti2012}.

In this article, we present a scientific case in the area of applicability of the GMCAO technique. Some preliminary studies were performed, recovering the structural parameters of a sample of synthetic nearby galaxies observed with a GMCAO-assisted ELT \citep{Portaluri2016a} and performing the source detection of a sample of high-redshift galaxies \citep{Portaluri2016b}.
These studies give a hint that the GMCAO approach can produce robust results when studying the photometry of extragalactic fields and can provide a useful frame of reference for a number of science cases. Here we want to extend those results by performing a more detailed analysis.

We took the CDFS as a reference for a star-poor field, which is the key (and a mandatory condition) for carrying out deep surveys, and we analyse a close-to-real observation in that region with a plausible giant telescope.
With this aim, we developed a tomographic simulator, which corrects the perturbation of wavefronts due to the atmosphere, and compute the Strehl ratio (SR) map measurements, as explained in Section~\ref{sec:simtool}.
Section~\ref{sec:PSF} describes how we built the point spread function (PSF), starting from the instrumental setup and the tomographic results. A description of how we obtained a mock image of a CDFS-like observation is given in Section~\ref{sec:field}. The source detection and the photometric analysis are shown and discussed in Section~\ref{sec:ana}. The conclusions of our work are presented in Section~\ref{sec:concl}.

\section{The Simulation of GMCAO Performance in the CDFS region}
\label{sec:simtool}
We used an Interactive Data Language (IDL) simulation tool \citep{Viotto2014} to estimate the sky coverage for a GMCAO system on an ELT and compute the resulting SR in a given FoV for a varying number of NGSs. Different technical FoV diameters for the references selection and various constellations were compared. 
\subsection{GMCAO Input Parameters}
The SR performance depends on a number of input parameters, \ie\ on both the assumed atmospheric behaviour and the simulated system layout. 
The atmospheric model is composed of 40 turbulent screens, based on the 40-layers $C_n^2$ profile \citep{Bergomi2014}, also used in \citet{Sarazin2013} and \citet{Auzinger2015}. The projected thickness (for zenith distance of $30^{\circ}$) of the atmosphere in this model is 29.1 km, the Fried parameter at 500 nm is $r_0=0.129$ m and the outer scale is $L_0$=25 m. Recall that the input $C_n^2$ profile plays a crucial role in the system performance estimation, especially for high altitude layers. This is true for all AO simulations, but it is particularly relevant in this case, because of the wider technical FoV involved and the higher sensitivity to high altitude layers. Some a priori knowledge of the atmosphere (not assumed in here), moreover, would certainly lead us to obtain a fully optimized performance.

The telescope aperture is assumed to be a non-obstructed 37 m circular pupil. The AO system includes three deformable mirrors conjugated to 0, 4, and 12.7 km, respectively, and assumes six GMCAO virtual deformable mirrors \citep{Ragazzoni2010}.
For each FoV we took into account up to six guide stars in a circular 10 arcmin technical FoV with a magnitude limit up to $R$ = 18 mag, discarding those within the inner 2 arcmin and those with a separation lower than 10 arcsec (for practical reasons). This choice leads to a GMCAO approach compatible with an ELT-like system equipped with MCAO.
\subsection{Sky Coverage Challenge: the {\it Chandra} Deep Field South}
\begin{figure*}
  \centering
  \begin{tabular}{c}
    \includegraphics[width=0.98\hsize,angle=0]{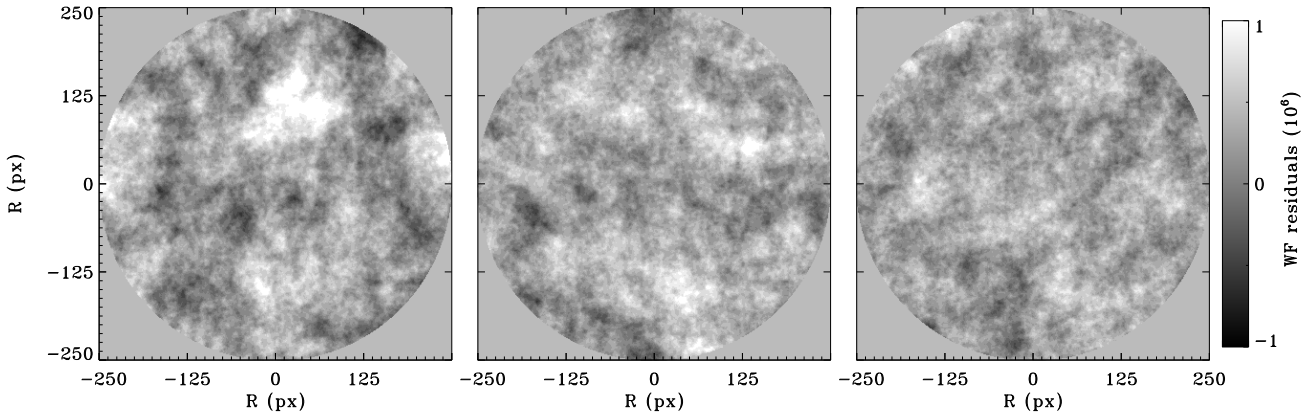}
  \end{tabular}  
\caption{\small{3 examples of wavefront residuals for 3 different SR obtained with the tomographic simulation tool: 0.05 (left-hand panel), 0.15 (central panel), 0.25 (right-hand panel). The scale is the same in all the cases.}}
\label{fig:WF}
\end{figure*}

\begin{figure*}
  \centering
    \begin{tabular}{c}
      \includegraphics[width=0.98\hsize,angle=0]{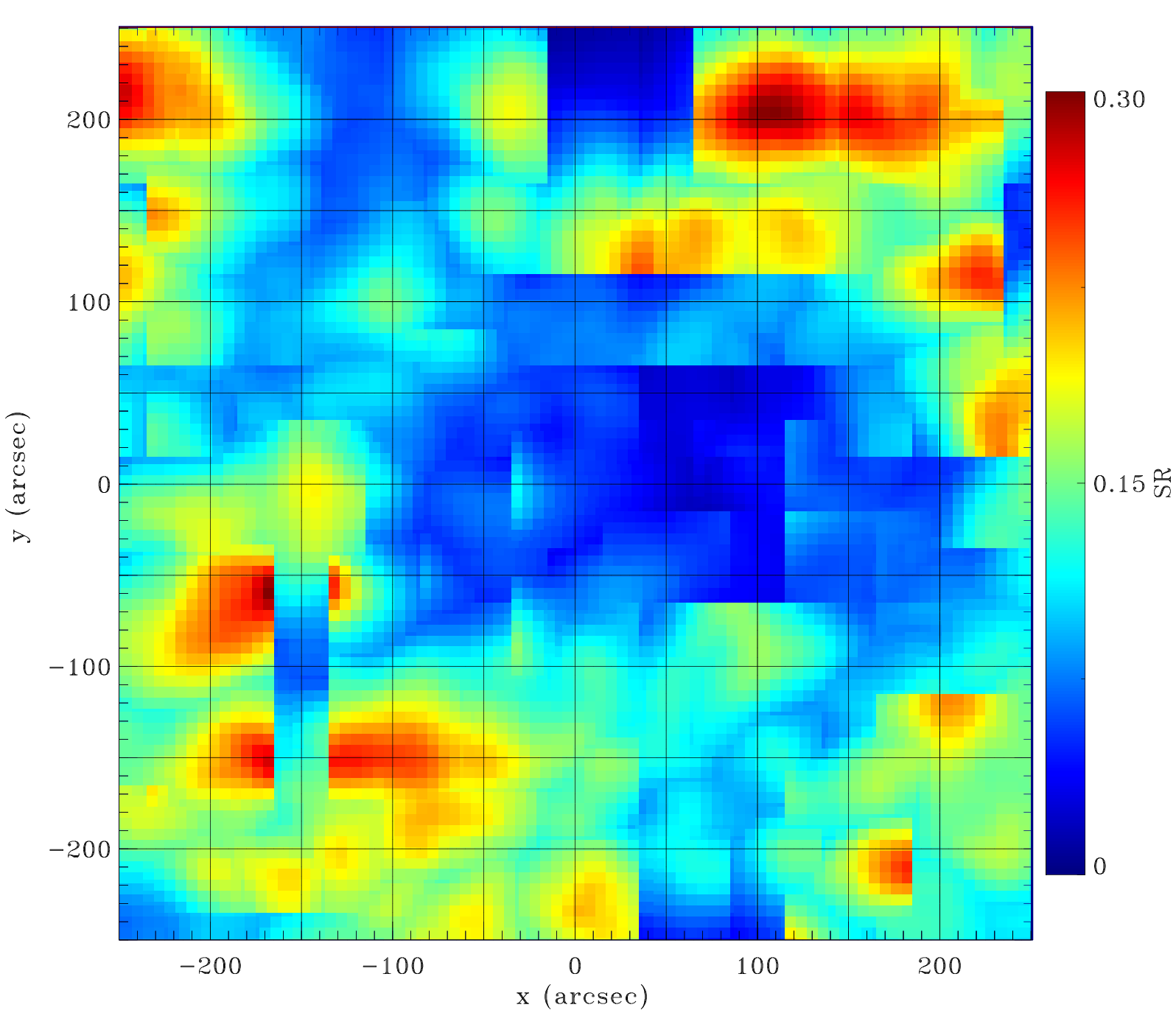}
    \end{tabular}
\caption{\small{$K$-band SR map obtained with the tomographic simulation tool pointing at the CDFS. The image is divided in 100 sectors of 50 $\times$ 50 arcsec$^2$ each. As discussed in the text, the strong discontinuities are due to the dithering technique that we adopted to build such a map: we selected the best value of the SR for each sector; but in this way adjacent regions could have a rich or poor asterism whenever one of the brighter and closer to the scientific FoV star is lost.}}
\label{fig:SRmap}
\end{figure*}
In order to assess the system's sky coverage, we used the USNO-B1.0 catalogue to retrieve $R$-band magnitude and astrometry of stars around the central coordinates of the CDFS ($\alpha=03^h32^m28^s$ and $\delta=-27^{\circ}48'30"$).
We considered a grid of 10$\times$10 technical FoVs each of 10 arcmin diameter, with a 50 arcsec pitch. 
For each FoV, we selected the asterism leading to the highest SR in the 50$\times$50 arcsec$^2$ scientific FoV (called ``sectors" hereafter), computed from the wavefront residuals using the formula (\citealt{Marechal1947} approximation) 
\begin{equation}
SR=\exp{(-\sigma^2)}
\end{equation}
where $\sigma$ is the root-mean-square (RMS) wavefront error in radians at the observing wavelength ($\lambda=2.2$ $\mu$m).
Three examples of wavefront residuals obtained from the tomographic simulation tool are shown in Figure~\ref{fig:WF}. They were used for building the PSFs of the corresponding SRs: 0.05, 0.15, and 0.25. As expected, the latter is more homogeneous with respect to the one with low SR.

\begin{figure*}
  \centering
  \begin{tabular}{c}    
    \includegraphics[width=0.98\hsize,angle=0]{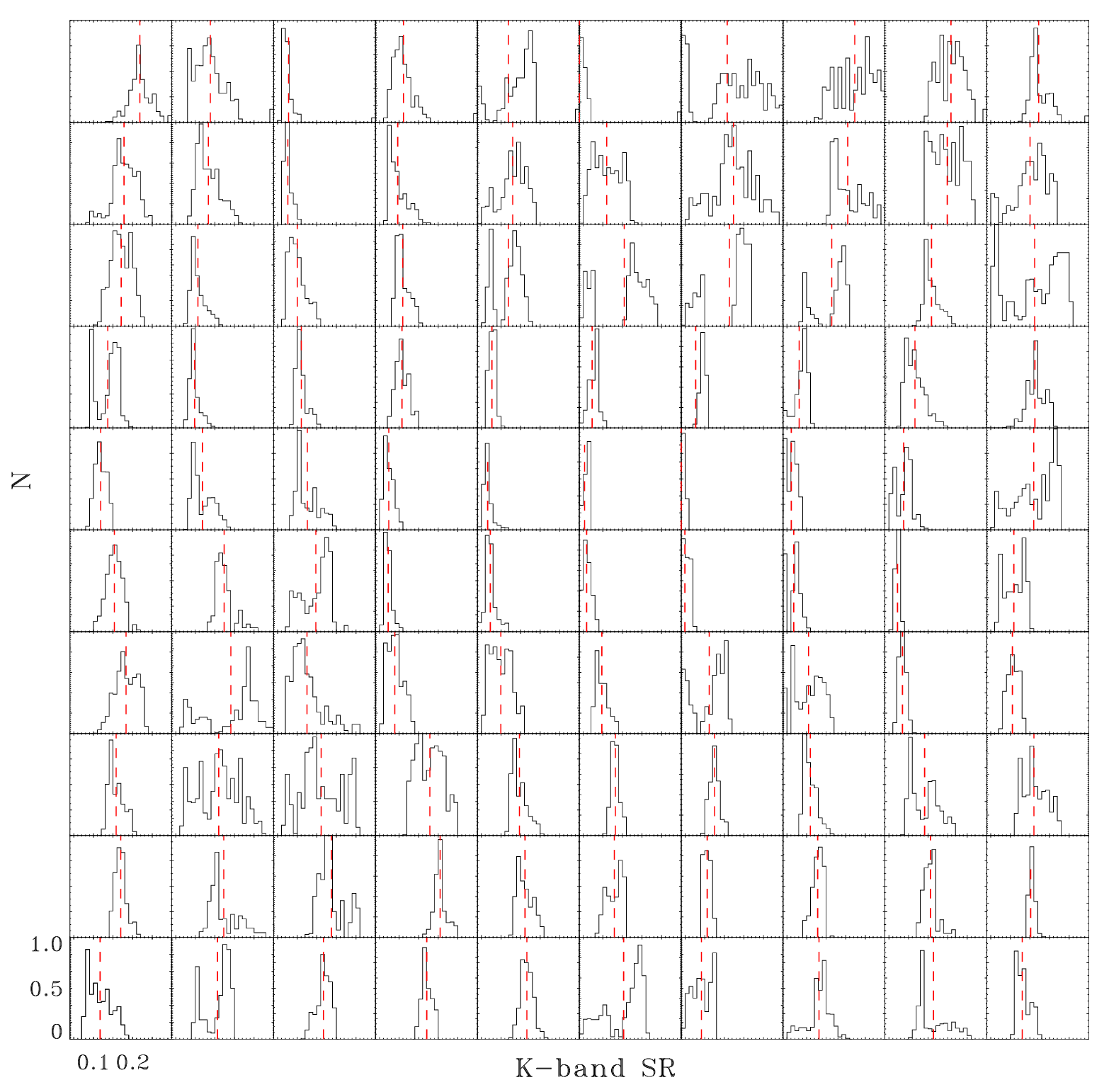}
  \end{tabular}  
\caption{\small{Histograms of the normalized distribution for the 100 sectors of 50 $\times$ 50 arcsec$^2$ each in the CDFS. The abscissae range, representing the $K$-band SR values, is the same for each plot: [0.05,0.3]. The dashed red dashed lines represent the mean value of the sector, reported in the map of the averaged SRs of Figure~\ref{fig:SRscacchi} (left-hand panel).}}
\label{fig:SRhist}
\end{figure*}

\begin{figure*}
  \centering
  \begin{tabular}{cc}    
\includegraphics[width=0.48\hsize,angle=0]{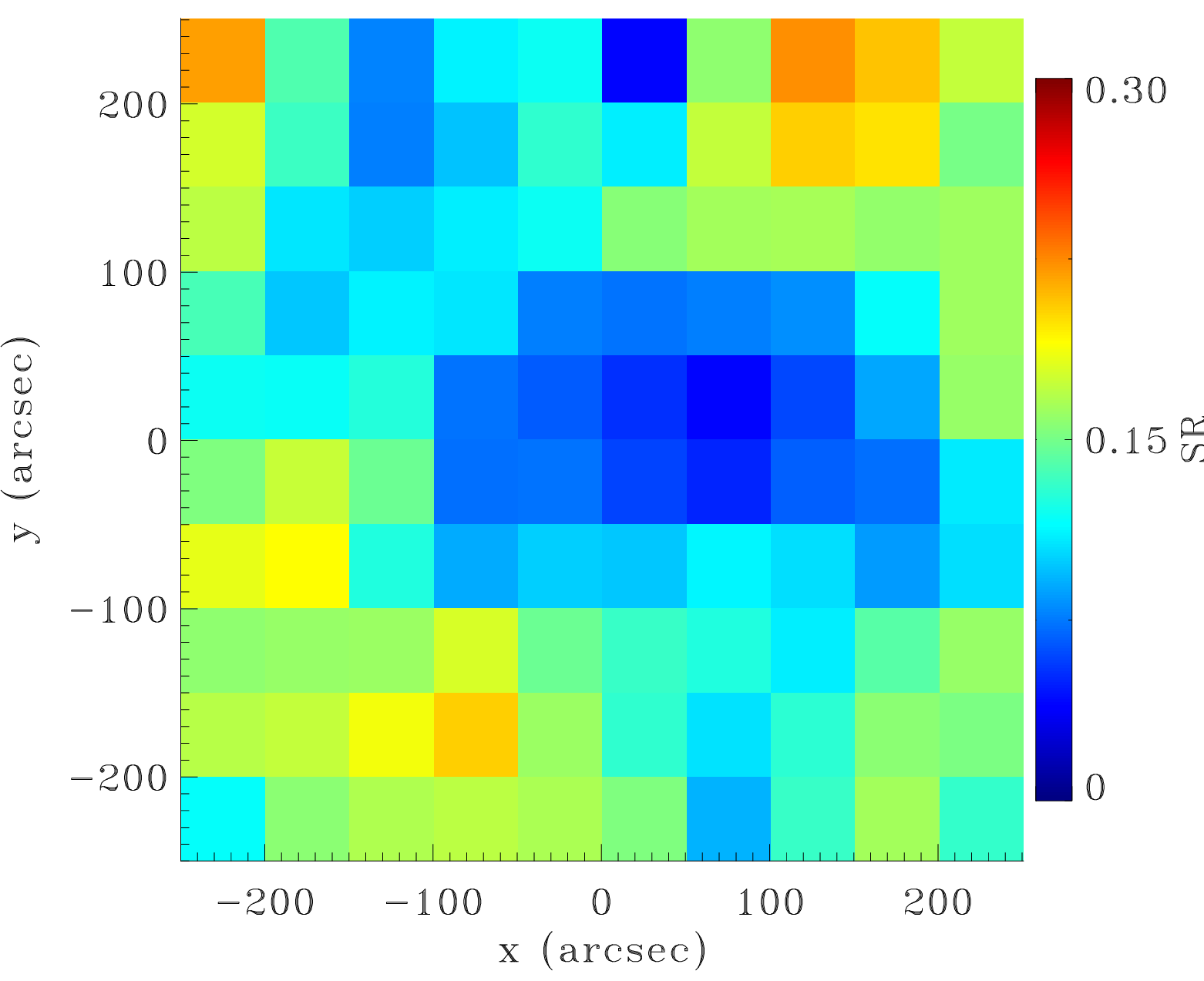}
\includegraphics[width=0.48\hsize,angle=0]{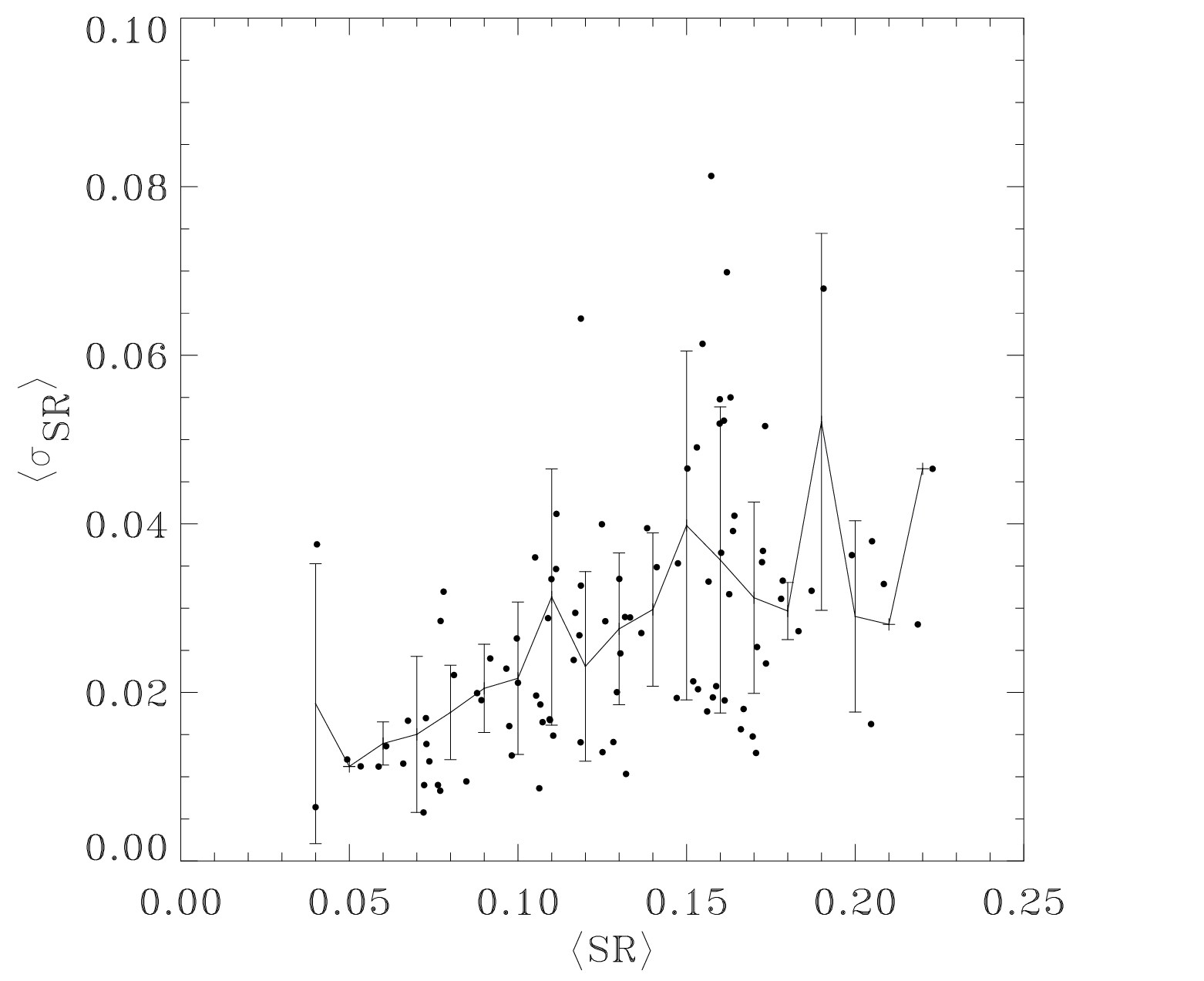}
  \end{tabular}
\caption{\small{Left-hand panel: $K$-band SR map of the average values measured within 100 sectors of 50 $\times$ 50 arcsec$^2$ each. Right-hand panel:
    The black points represent $\left < \sigma_{\rm SR} \right >$ for each sector as a function of their mean SR. The black line shows the mean of the values within the SR bin, and 1$\sigma$ error is indicated by the error bars.}}
\label{fig:SRscacchi}
\end{figure*}

Figure~\ref{fig:SRmap} shows the $K$-band SR map we obtained for the FoVs, covering a total region of $500 \times 500$ arcsec$^2$, subdivided, in 10$\times$10 sectors whose size are of the same order as the largest typical FoV planned for next generation ELTs.
The SR values obtained range between 0.01 and 0.29, and the mean is 0.17.
The discontinuities that are present in some regions of the map are due to the dithering technique we adopted.
In fact, in some cases, the selected asterism is good enough to let us reach a high SR (\ie\ the stars were brighter and closer to the scientific FoV, but outside the inner 2 arcmin), while in the contiguous field we miss one or more important guide stars and therefore the SR gets lower.
 For this reason, the distribution of SR values inside each sector varies significantly, as shown in Figure~\ref{fig:SRhist} and can not be associated to a unique analytical distribution. As expected, the sectors with high discontinuity have a higher spread.
The mean SRs are indicated by the red dashed line: some sectors show the same average value (spanning between 0.4 and 0.22), but have a different spread.
This is even more clear in Figure~\ref{fig:SRscacchi}, where a 2-dimensional map of the mean values is shown. The central regions of the map have lower values of SR, mainly due to the fact that there are few guide stars available, while in the corners the SR becomes higher. The distribution of $<\sigma_{\rm SR}>$ for each sector as a function of its mean SR shows that there is a clear trend: low SRs tend to have low standard deviations, while sectors with higher SRs have a wider distribution.
This finding is not surprising: in fact, given the irregular and random nature of the NGSs positions, peak performances are occasionally obtained in some portions of the FoV, thus leading to a larger scatter in the SR distribution. However, such an effect is expected to be less relevant in GMCAO than in a conventional MCAO approach, for the following twofold reason \citep{Ragazzoni2010}:
  \begin{enumerate}
  \item the randomness of NGSs (brightness and position) is less severe in a statistical sense, because of the larger FoV;
  \item the angle of arrival of the NGSs rays is larger than in a MCAO approach and therefore the wavefront residuals, which are responsible for the spatial dependence of PSF quality, are less effective. This is due to the fact that the spreading of the sensed turbulence away from the deformable mirrors is larger.
  \end{enumerate}


\section{The GMCAO PSF Construction}
\label{sec:PSF}

\begin{table}
\caption{\small{Weight parameters for building the final PSF.}}
\label{tab:PSF}
\begin{scriptsize}
\begin{center}
\begin{tabular}{ccccc}
\hline
\multicolumn{1}{c}{SR} &
\multicolumn{1}{c}{$\eta$} &
\multicolumn{1}{c}{$h_{GAUSS}$} &
\multicolumn{1}{c}{$\frac{h_{TOT}}{h_{DL}}$} \\
\multicolumn{1}{c}{} &
\multicolumn{1}{c}{} &
\multicolumn{1}{c}{$\times 10^{-6}$ [A.U.]} &
\multicolumn{1}{c}{} \\
\hline
  0.01  &      0.009 &  1.923 &  0.01  \\
  0.02  &      0.018 &  1.912 &  0.02  \\
  0.03  &      0.032 &  1.906 &  0.03  \\
  0.04  &      0.043 &  1.895 &  0.04  \\
  0.05  &      0.052 &  1.882 &  0.05  \\
  0.05$^*$  &0.054 &  1.886 &  0.05 \\
   0.05$^*$ &0.066 &  1.900 &  0.05\\
   0.05$^*$ &0.053 &  1.885 &  0.05\\
   0.05$^*$ &0.052 &  1.882 &  0.05\\
   0.05$^*$ &0.055 &  1.886 &  0.05\\     
  0.06  &      0.076 &  1.888 &  0.06  \\
  0.07  &      0.076 &  1.865 &  0.07  \\
  0.08  &      0.092 &  1.857 &  0.08  \\
  0.09  &      0.114 &  1.863 &  0.09  \\
  0.10  &      0.129 &  1.857 &  0.10  \\
  0.11  &      0.122 &  1.828 &  0.11  \\
  0.12  &      0.191 &  1.883 &  0.12  \\
  0.13  &      0.185 &  1.853 &  0.13  \\
  0.14  &      0.172 &  1.809 &  0.14  \\
  0.15  &      0.181 &  1.792 &  0.15  \\
   0.15$^*$  &0.265 &  1.919 &  0.15\\
   0.15$^*$ &0.217 &  1.850 &  0.15\\
   0.15$^*$ &0.251 &  1.902 &  0.15\\
   0.15$^*$ &0.219 &  1.846 &  0.15\\
   0.15$^*$ &0.176 &  1.786 &  0.15\\
  0.16  &      0.202 &  1.800 &  0.16  \\
  0.17  &      0.189 &  1.757 &  0.17  \\
  0.18  &      0.256 &  1.816 &  0.18  \\
  0.19  &      0.268 &  1.793 &  0.19  \\
  0.20  &      0.239 &  1.726 &  0.20  \\
  0.21  &      0.228 &  1.697 &  0.21  \\
  0.22  &      0.273 &  1.710 &  0.22  \\
  0.23  &      0.323 &  1.753 &  0.23  \\
  0.24  &      0.368 &  1.793 &  0.24  \\
  0.25  &      0.491 &  1.849 &  0.25  \\
   0.25$^*$ &0.428 &  1.897 &  0.25\\
   0.25$^*$ &0.373 &  1.789 &  0.25\\
   0.25$^*$ &0.491 &  1.749 &  0.25\\
   0.25$^*$ &0.396 &  1.800 &  0.25\\
   0.25$^*$ &0.320 &  1.692 &  0.25\\
  0.26  &      0.404 &  1.779 &  0.26  \\
  0.27  &      0.435 &  1.801 &  0.27  \\
  0.28  &      0.349 &  1.630 &  0.28  \\
  0.29  &      0.366 &  1.625 &  0.29  \\
 \hline
\end{tabular}
\end{center}
\begin{minipage}{8.5cm}
NOTES -- Col.(1): SR bins of the PSFs. SRs labelled with $^*$ represent the control sample used for testing the dependence of the final PSF on the selected wavefront
Col.(2): Weight parameters adopted to sum the diffraction-limited PSF and the Gaussian PSF in order to obtain the final PSF, and
Col.(3): Maximum of the Gaussian PSF, as resulted from the solution of the system in Equation~\ref{eq:PSFsystem}.
Col.(4): Ratio between the peak of the final and the diffraction-limited PSF, which should (and does) correspond to the SR.
\end{minipage}
\label{minipage_tab}
\end{scriptsize}
\end{table}
The convolution of the surface-brightness distribution with the PSF is of paramount importance for obtaining close-to-real images and therefore simulating an observation properly.
We built 29 $K$-band synthetic PSFs assuming a 0.1-wide bin, starting from 0.01 to 0.29. The core shape is given by the wavefront matrix error, which is the output of the tomographic simulation tool and represents the residual map between the corrected and the flat (diffraction limited) wavefronts. A two-dimensional Gaussian function was then added to represent the external halo. The procedure is described fully in \citet{Portaluri2016b}.
The final PSF ($PSF_{TOT}$) is obtained by weighting the two components ($PSF_{wavef,i}$ and $PSF_{Gauss,i}$) properly according to the corresponding bin ($i$) of SR:
\begin{equation}
  \label{eq:PSFsystem}
\begin{cases}
  PSF_{TOT,i}= \eta \cdot PSF_{wavef,i}+(1-\eta) \cdot PSF_{GAUSS,i}
  \vspace{0.005cm}\\
  \frac{h_{TOT,i}}{h_{DL,i}}= SR_{i}
\end{cases}
\end{equation}
where $h_{TOT,i}$ and $h_{DL,i}$ are the maxima of the total and diffraction limited PSFs, in the $i^{\rm th}$-bin, respectively.  $\eta < 1$ is the weight parameter, which is an unknown variable, together with $h_{GAUSS,i}$, which is the maximum of the Gaussian function necessary to determine $h_{TOT,i}$.
The results of the equation system for our observations are listed in Table~\ref{tab:PSF}: the higher the SR, the smaller the Gaussian contribution.

To measure how much the final PSFs depend on the particular selected wavefronts, we considered a control sample of five synthetic wavefronts for the SR bins equal to 0.05, 0.15, and 0.25. Thus, we built five PSFs from the different wavefront error matrices, using the values listed in Table~\ref{tab:PSF} and labelled with $^*$. 
  We averaged the obtained PSFs of each bin ($PSF_{\rm ctr,i}$) and calculated the relative residual images using the previously built PSFs as a reference ($PSF_{\rm ref,i}$):
  \begin{equation}
      \label{eq:PSFtest}
\epsilon_{PSF,i}= \left < {\frac{PSF_{\rm ref,i}- PSF_{\rm ctr,i}}{PSF_{\rm ref,i}}} \right >
\end{equation}
  The results show that $\epsilon_{PSF}= 5$ per cent, $10$ per cent, and $13$ per cent for the 0.05, 0.015, and 0.25 bin, respectively. The highest source of errors is given by the wings, which contain most of the noise. As expected, the error grows when considering higher SR because the PSF becomes highly dependent on the $PSF_{wavef,i}$ and the contribution of the Gaussian function is lower. The net conclusion is that the kernel of the PSF is well described, while the wings can be affected by an error of the order of 10 per cent. This finding allows for future studies of a certain kind of science, in which the central region of the PSF plays a key role. However, these fluctuations are not significant for the rest of the analysis: we have verified that, for several realizations of the PSF within the same bin of SR, the standard deviation of the energy enclosed in the core is $\approx 1$ per cent, which translates into an equivalent variation of SR up to $0.2$ per cent. These differences lead to errors in the analysis that are significantly smaller than the ones caused by a wrong PSF reconstruction, which are taken into account in this work and discussed in Section~\ref{sec:ana}, and hence neglected in the following considerations. 

\section{The Simulation of a Deep Extragalactic Field in the CDFS Region}
\label{sec:field}
To address the imaging capabilities of a GMCAO-based system mounted on an ELT, we need to simulate a typical field with characteristics adequate to the ELTs instrumentation and properties.
We used a 10$\times$10 grid of $50\times50$ arcsec$^2$ FoV, matching the largest typical FoV planned for next generation of telescopes \citep{Wright2014,Sharp2014,Davies2016}.
Each image was obtained by introducing simple models of astronomical objects, \ie\ stars and galaxies, as specified in an input list, using the public software AETC\footnote{http://aetc.oapd.inaf.it/} v.3.0 \citep{Falomo2011}. 
The procedure took into account the effects of the pixel scale (selected in order to be compatible with the planned next-generation imagers, \ie\ 0.003 arcsec px$^{-1}$), pixel sampling, background contribution, gain, readout and Poisson noise, and was already used with a similar purpose by \citet{Gullieuszik2016}. The selected near-IR sky background was the one measured at Cerro Paranal (12.8 mag arcsec$^{-2}$ in the $K$ band) and included the contribution of thermal emission in the near-IR-bands. The integration time used was 3 hours.
The choice of the $K$ band is explained by the fact that the best astronomical performances for ELTs are expected in the near-IR since they will make use of AO techniques. In fact, as the Fried scalelength depends on $\lambda ^{6/5}$ and the corresponding parameters (Greenwood frequency and isoplanatic patch) scale accordingly, AO is notoriously less well-performing toward the bluer part of the spectrum \citep{Davies2012}.
The profiles and templates are given elliptical shapes by specifying the scale radius for the major axis, $R_{\rm e}$, the ellipticity, ${\rm ell}= 1- b/a$, where $b/a$ is the minor to major axis ratio, the position angle, $PA$, defined counterclockwise from the $x$ axis, and the coordinates of the centre, $(x_0,y_0)$, of the object.
The objects are also specified by the photometric laws that connotate their morphological types and their total magnitude, $K_{tot}$.
We used simple flux profiles template, defined by the S\'ersic law:
\begin{equation}
  \mu(r)=\mu_e + 2.5 \frac{b_n}{\ln(10)} \left[ \left( \frac{r}{R_{\rm e}} \right ) ^{1/n}-1 \right ]
\end{equation}
\begin{equation}
  \mu_e=K_{tot}+ 2.5 \log[(1-e) 2\pi R_{\rm e}^2]+ 2.5 \log \left [ n \rm{e}^{b_n} \frac{\Gamma(2n)}{b_{n}^{2n}}\right ]
\end{equation}
where $\mu$ is the surface brightness of the galaxy, $e$ the ellipticity, $\Gamma$ the complete gamma function.
The S\'ersic index $n$ is linked to the morphology: in particular for $n=1$ the distribution becomes exponential and represents discy galaxies \citep{Freeman1970}, while a profile with $n=4$ is common for elliptical galaxies \citep{DeVaucouleurs1992}.
$R_{\rm e}$ is the effective radius, which encloses half the flux, often used as measure of the size of a galaxy, while $\mu_{\rm e}$ is the magnitude value at $R_{\rm e}$.

As our main goal is to test the performance of a GMCAO-based system in recovering the global structural parameters of (high-redshift) galaxies (magnitude, size and morphology) we did not take into account more complicated models or galactic substructures.
\subsection{Stars Input Parameters}
When analysing real images, it is crucial to first deconvolve the surface-brightness distribution from the effects of the PSF in order to derive the photometric parameters of galaxies properly: we needed some stars to build the instrumental PSF for the following analysis. 
Considering a plausible limit that the next-generation telescopes will be able to reach (\ie\ $K$=30 mag in the AB magnitude system), we should find a mean of 3 stars in a 50$\times$50 arcsec$^2$ field. To obtain this number we investigated 100 fields of $50\times50$ arcsec$^2$ each, around the CDFS coordinates, using TRILEGAL \citep{Girardi2005}, a population synthesis code, which simulates the stellar photometry of our Galaxy, and averaged the results.
 Therefore, we placed 3 stars with random positions in each sector and, averaging and normalizing them, we obtained 100 experimental PSFs (1 per sector) to be used for the deconvolution.

\subsection{Galaxies Input Parameters}
\label{subsec:input}
\begin{table}
\caption{\small{Input parameters for the galaxies simulation of early- ($n$=4) and late-type galaxies ($n$=1).}}
\label{tab:input}
\begin{scriptsize}
\begin{center}
\begin{tabular}{crrrrrr} 
\hline
\multicolumn{1}{c}{Obj. ID\#} &
\multicolumn{1}{c}{$\log {M/M_{\odot}}$} &
\multicolumn{1}{c}{ $z$} &
\multicolumn{1}{c}{$K_{tot}$} &
\multicolumn{1}{c}{$R_{\rm e}$} &
\multicolumn{1}{c}{$b/a$} &
\multicolumn{1}{c}{$PA$}\\
\multicolumn{1}{c}{} &
\multicolumn{1}{c}{} &
\multicolumn{1}{c}{} &
\multicolumn{1}{c}{[mag]} &
\multicolumn{1}{c}{[arcsec]} &
\multicolumn{1}{c}{} &
\multicolumn{1}{c}{[\degrees]} \\
\hline
  &       &         &    $n$=4 &            &         &              \\
1 &  9.0  &    0.25 & 21.67  &    0.052   &  0.372  &    304.1       \\
2 &  9.0  &    0.75 & 21.00  &    0.087   &  0.581  &    9.9         \\
3 &  9.0  &    1.25 & 20.11  &    0.175   &  0.178  &    104.6       \\
4 &  9.0  &    1.75 & 19.44  &    0.294   &  0.346  &    354.1       \\
5 &  9.0  &    2.25 & 18.77  &    0.494   &  0.893  &    118.7       \\
6 &  9.0  &    2.75 & 23.42  &    0.027   &  0.495  &    47.0        \\
7 &  9.3  &    0.25 & 22.71  &    0.044   &  0.861  &    52.5        \\
8 &  9.3  &    0.75 & 21.75  &    0.083   &  0.949  &    230.8       \\
9 &  9.3  &    1.25 & 21.04  &    0.132   &  0.706  &    242.9       \\
10&  9.3  &    1.75 & 20.33  &    0.212   &  0.610  &    206.9       \\
11&  9.3  &    2.25 & 24.48  &    0.009   &  0.449  &    324.9       \\
12&  9.3  &    2.75 & 23.83  &    0.016   &  0.772  &    341.8       \\
13&  9.7  &    0.25 & 22.97  &    0.032   &  0.900  &    303.0       \\
14&  9.7  &    0.75 & 22.32  &    0.055   &  0.699  &    99.2        \\
15&  9.7  &    1.25 & 21.68  &    0.095   &  0.550  &    167.6       \\
16&  9.7  &    1.75 & 24.42  &    0.008   &  0.732  &    328.7       \\
17&  9.7  &    2.25 & 23.81  &    0.013   &  0.654  &    218.3       \\
18&  9.7  &    2.75 & 23.00  &    0.026   &  0.404  &    325.6       \\
19&  10.0  &   0.25 & 22.39  &    0.044   &  0.412  &    232.1       \\
20&  10.0  &   0.75 & 21.78  &    0.074   &  0.452  &    127.0       \\
21&  10.0  &   1.25 & 24.46  &    0.006   &  0.517  &    162.2       \\
22&  10.0  &   1.75 & 23.90  &    0.009   &  0.696  &    302.5       \\
23&  10.0  &   2.25 & 23.14  &    0.019   &  0.675  &    128.3       \\
24&  10.0  &   2.75 & 22.58  &    0.032   &  0.990  &    168.9       \\
25&  10.3  &   0.25 & 22.01  &    0.054   &  0.638  &    262.4       \\
26&  10.3  &   0.75 & 25.35  &    0.005   &  0.174  &    33.5        \\
27&  10.3  &   1.25 & 24.82  &    0.009   &  0.816  &    174.5       \\
28&  10.3  &   1.75 & 24.11  &    0.019   &  0.303  &    267.4       \\
29&  10.3  &   2.25 & 23.57  &    0.032   &  0.988  &    19.5        \\
30&  10.3  &   2.75 & 23.04  &    0.054   &  0.996  &    112.7       \\
\hline
&       &         &$n=$1     &            &         &               \\
31 &9.0   &    0.25 & 20.46  &    0.663   &  0.250  &    214.5      \\
32 &9.0   &    0.75 & 19.71  &    0.789   &  0.745  &    289.6      \\
33 &9.0   &    1.25 & 18.71  &    0.995   &  0.714  &    64.20      \\
34 &9.0   &    1.75 & 17.96  &    1.184   &  0.846  &    298.6      \\
35 &9.0   &    2.25 & 17.21  &    1.408   &  0.100  &    85.5       \\
36 &9.0   &    2.75 & 21.87  &    0.388   &  0.990  &    131.8      \\
37 &9.3   &    0.25 & 21.16  &    0.450   &  0.389  &    158.5      \\
38 &9.3   &    0.75 & 20.23  &    0.549   &  0.654  &    127.0      \\
39 &9.3   &    1.25 & 19.53  &    0.637   &  0.532  &    27.2       \\
40 &9.3   &    1.75 & 18.83  &    0.739   &  0.435  &    291.5      \\
41 &9.3   &    2.25 & 22.79  &    0.255   &  0.487  &    273.9      \\
42 &9.3   &    2.75 & 22.20  &    0.296   &  0.478  &    69.4       \\
43 &9.7   &    0.25 & 21.41  &    0.360   &  0.988  &    194.8      \\
44 &9.7   &    0.75 & 20.82  &    0.418   &  0.343  &    309.4      \\
45 &9.7   &    1.25 & 20.23  &    0.484   &  0.984  &    233.6      \\
46 &9.7   &    1.75 & 23.63  &    0.220   &  0.793  &    39.9       \\
47 &9.7   &    2.25 & 23.03  &    0.257   &  0.978  &    12.0       \\
48 &9.7   &    2.75 & 22.24  &    0.317   &  0.410  &    151.3      \\
49 &10.0   &   0.25 & 21.65  &    0.371   &  0.113  &    43.2       \\
50 &10.0   &   0.75 & 21.05  &    0.434   &  0.864  &    29.5       \\
51 &10.0   &   1.25 & 24.03  &    0.179   &  0.841  &    299.5      \\
52 &10.0   &   1.75 & 23.44  &    0.208   &  0.190  &    155.8      \\
53 &10.0   &   2.25 & 22.64  &    0.256   &  0.510  &    62.7       \\
54 &10.0   &   2.75 & 22.04  &    0.298   &  0.532  &    54.4       \\
55 &10.3   &   0.25 & 21.45  &    0.348   &  0.906  &    300.8      \\
56 &10.3   &   0.75 & 24.22  &    0.202   &  0.877  &    49.3       \\
57 &10.3   &   1.25 & 23.68  &    0.228   &  0.985  &    171.4      \\
58 &10.3   &   1.75 & 22.97  &    0.269   &  0.237  &    129.8      \\
59 &10.3   &   2.25 & 22.44  &    0.305   &  0.702  &    298.3      \\
60 &10.3   &   2.75 & 21.91  &    0.345   &  0.382  &    195.0      \\
\hline               
\end{tabular}
\end{center}
\begin{minipage}{8.5cm}
  NOTES --
  Col.(1): Simulated object ID number.
  Col.(2): Logarithm of the $M/M_{\odot}$.
  Col.(3): Redshift.
  Col.(4): $K$-band total magnitude.
  Col.(5): Effective radius for the galaxies.
  Col.(6): axial ratio, i.e. minor axis above major axis.
  Col.(7): Position angle with respect to the North.
\end{minipage}
\label{minpagetab:modelsim}
\end{scriptsize}
\end{table}
As we were dealing with deep observations, we considered a sample of 60 high-redshift galaxies (30 with $n=1$ and 30 with $n=4$) with random inclinations and PAs that represented our sample, as listed in Table~\ref{tab:input}.
They are observed at 6 different redshifts ($z= 0.25$, 0.75, 1.25, 1.75, 2.25, 2.75), and have a mass included in the list $\log {M/M_{\odot}}=9$,~$9.3$,~$9.7$,~$10$,~$10.3$, following the results of \citet{vanderWel2014}, who studied the relationship between the structural parameters (size, as defined by the effective radius, $R_{\rm e}$; luminosity, as measured from total $H$-band magnitude, $H$; and total mass, $M$) for 30000 galaxies from CANDELS HST survey. In this way, we pushed the observational limits of the CDFS $K$-band catalogue presented by \citet{Saracco2001}, obtained from VLT-ISAAC images centered on the CDFS, for a total area of 13.6 arcmin$^2$ with a limiting surface brightness of 22.8 mag arcsec$^{-2}$.
From the mass-size and the mass-luminosity relations, which are different for early- and late-type galaxies, we extrapolated the input parameters, and assumed a color of $H-K=1$ mag, as described in \citet{Portaluri2016b}. We used the concordance cosmology established by \cite{Planck2015}, for a flat Universe with $H_0=67.8$ km s$^{-1}$ Mpc$^{-1}$ and $\Omega_{\rm m}=0.308$. 

\begin{figure}
  \begin{center}
    \includegraphics[width=0.99\hsize,angle=0]{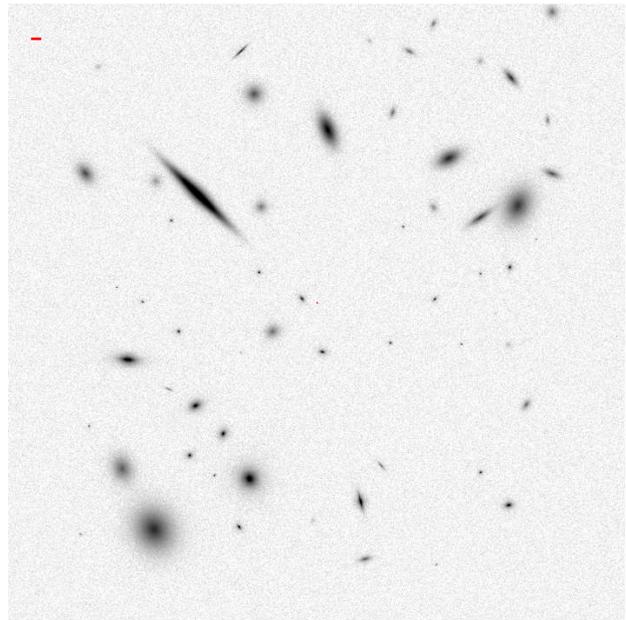}
\caption{\small{Mock $K$-band image of one sector of the deep field having 60 high-redshift galaxies as listed in Table~\ref{tab:input}. The FoV is 50$\times$50 arcsec$^2$. The red dash is 1 arcsec wide.}}
\label{fig:field}
  \end{center}
\end{figure}

All 60 objects were randomly placed in each sector, and Figure~\ref{fig:field} shows one realization. In this way we obtained a close-to-real observation of a 500$\times$500 arcsec$^2$ field containing 6000 galaxies. Each object in a given sector was different from its counterpart placed in another sector because it was convolved with a different PSF.
In fact, we used the random coordinates of the centre of each galaxy (and also of the 3 stars of the field) to find the corresponding value in the SR map, selecting the GMCAO PSF picked to perform the image convolution: therefore the 100 images of the sectors were different one from the other.

\section{Photometric Analysis}
\label{sec:ana}

\begin{figure*}
  \centering
  \begin{tabular}{c}
    \includegraphics[width=0.84\hsize,angle=0]{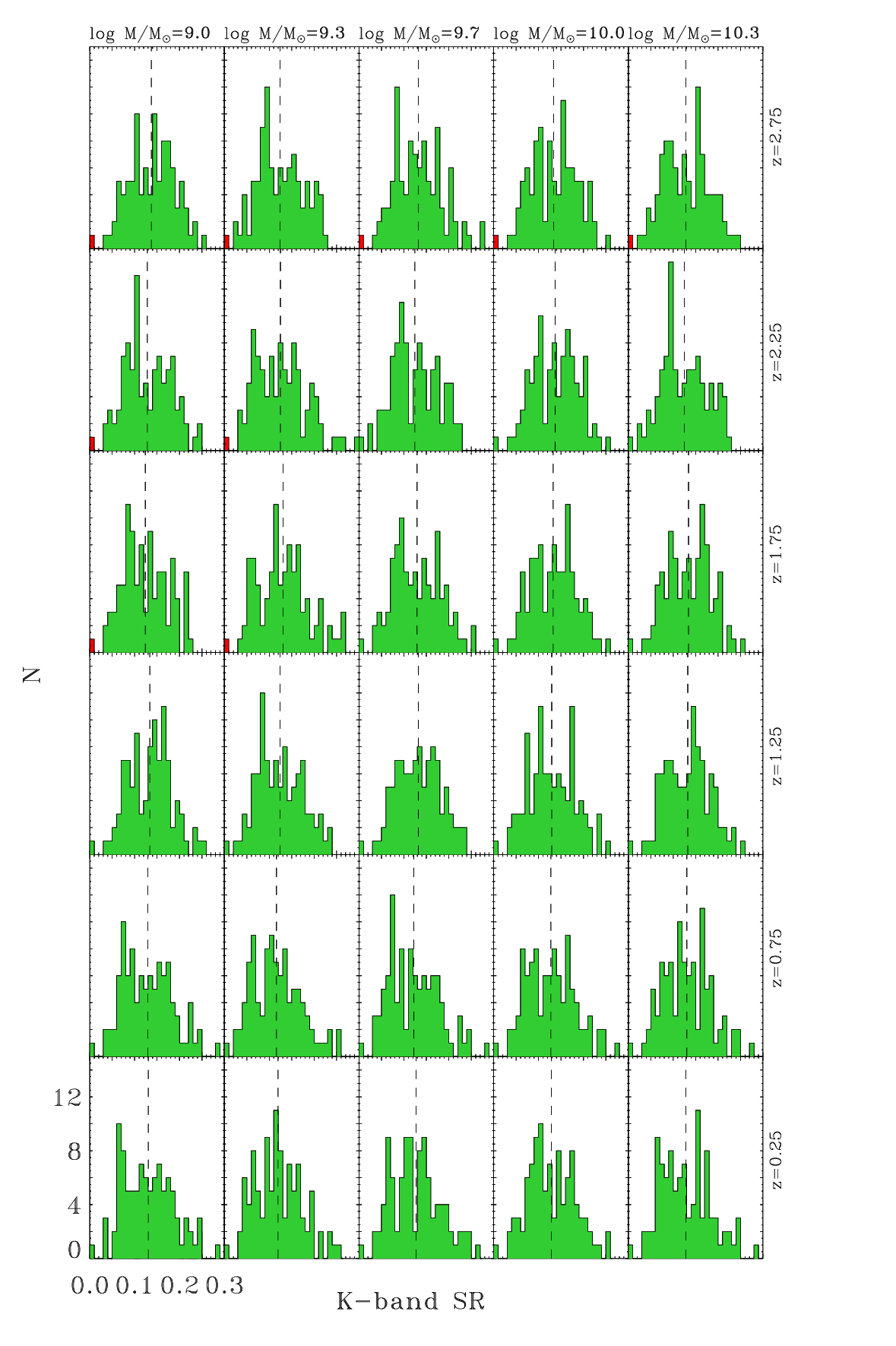}
  \end{tabular}
\caption{\small{Histograms of the number of objects detected as a function of the SR (and PSF) used to convolve the galaxy image. The analysis is shown for the 30 early-type galaxies of the sample contained in the 100 sectors. The objects are ordered by increasing mass from left to right and increasing redshift from bottom to top. Green regions show the cases where SExtractor has detected the object, while red regions are for missing detections. The vertical dotted lines indicate the mean values of the SRs for the 100 representations of each object.}}
\label{fig:sex}
\end{figure*}

\begin{figure*}
  \centering
  \begin{tabular}{c}    
    \includegraphics[width=0.84\hsize,angle=0]{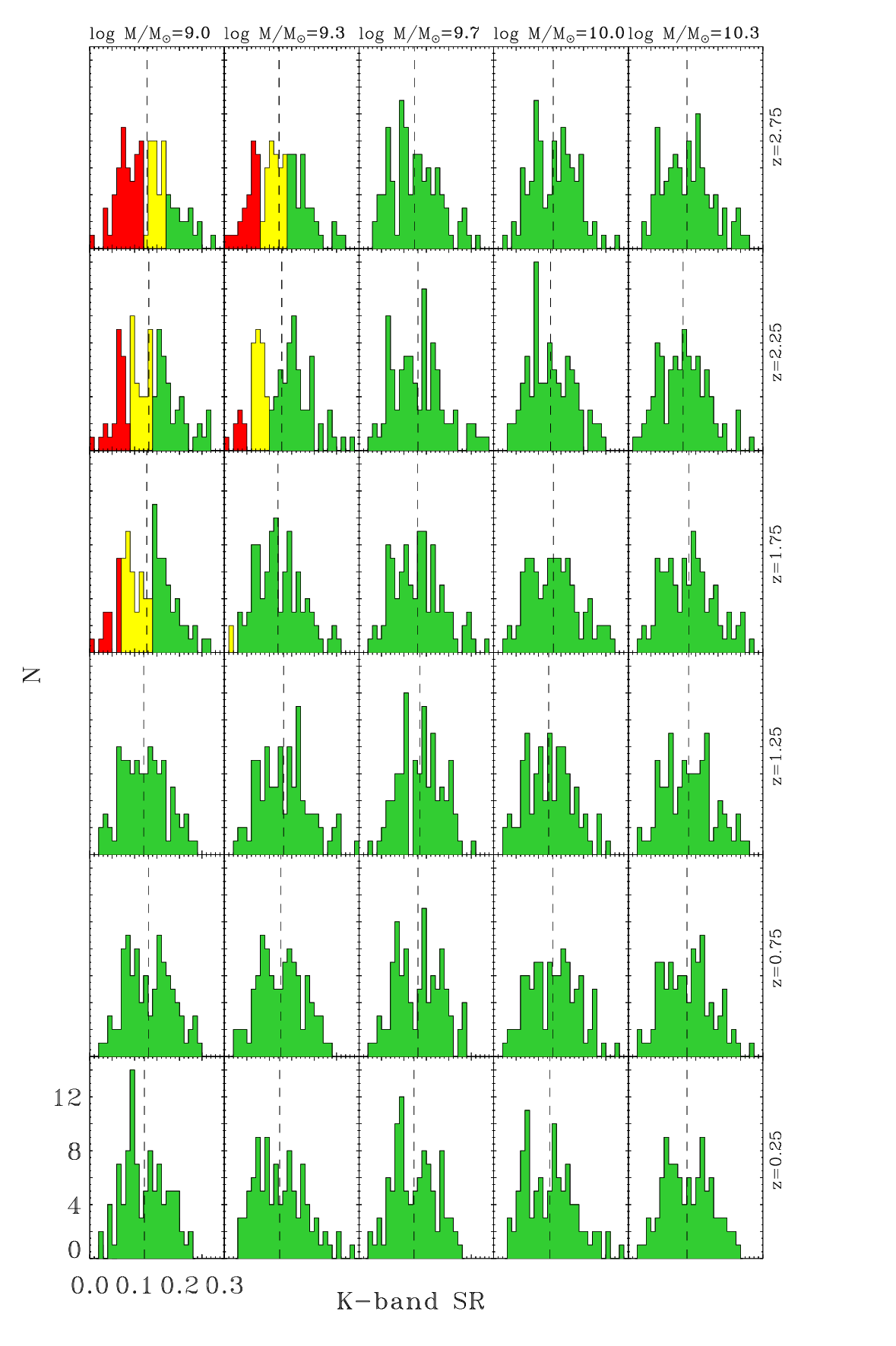}
  \end{tabular}
  \caption{\small{Histograms of the number of objects detected as a function of the SR (and PSF) used to convolve the galaxy image. The analysis is shown for the 30 late-type galaxies of the sample contained in the 100 sectors. The objects are ordered for increasing mass from left to right and increasing redshift from bottom to top. Green regions show the cases where SExtractor has detected the object, while red regions are for missing detections. We have also performed a 2$\times$2 rebinning, as explained in the text, to improve the S/N: this analysis increases the numbers of detected objects, and it is shown by the yellow regions. The vertical dotted lines indicate the mean values of the SRs for the 100 representations of each object.}}
\label{fig:sex2}
\end{figure*}

We performed source detection in all 100 sectors using SExtractor \citep{Bertin1996}, obtaining detection statistics as a function of its SR, as shown in Figure~\ref{fig:sex} and Figure~\ref{fig:sex2}. 
From left to right the objects in the figures increase in mass, while from bottom to top the redshift increases. The detection is done using $2\sigma$ above the background, a threshold level for all the runs. This value was chosen after a number of tests to detect the highest number of objects and minimise the number of spurious detections.
Green regions represent the cases where the detection was successful, while red regions indicate missing detections. We were able to easily detect 99.7 and 89.4 per cent of early- and late-type galaxies of our sample, respectively. The late-type objects have a fainter central surface brightness than their early-type counterparts (\ie\ galaxies that have the same total magnitude) and therefore the signal-to-noise ratio (S/N) is lower. To increase the S/N of the images with the late-type objects, we performed a $2\times2$ rebinning and performed the SExtractor estimation again, increasing to 95.3 per cent the source detection, as shown by the yellow regions of Figure~\ref{fig:sex2}.
The undetected objects are those at high redshift with low mass and SR.


\begin{figure}
\begin{center}
\includegraphics[width=0.98\hsize,angle=0]{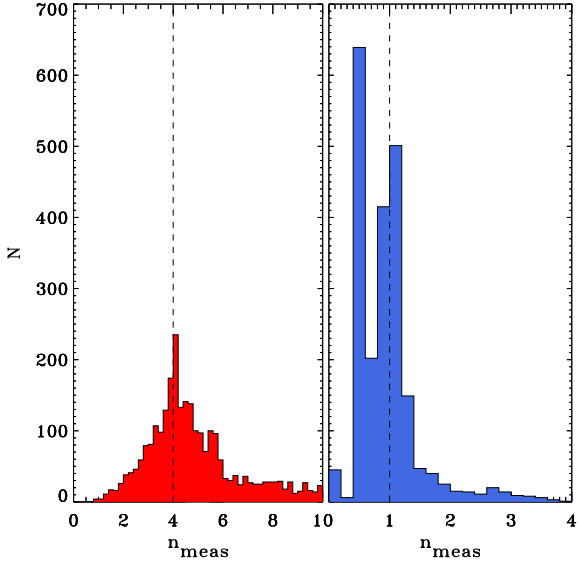}
\caption{\small{Histogram of the distribution of S\'ersic index for the early- (left-hand panel) and late-type (right-hand panel) galaxies of the sample. The dashed lines represent the value we expected according to their morphology.}}
\label{fig:nsers}
\end{center}
\end{figure}

The resulting surface-brightness distribution of each galaxy of each image was then modelled with a  S\'ersic component by applying the two-dimensional fitting algorithm GALFIT \citep{Peng2002} using the PSF obtained by averaging the three stars of the same sector for the deconvolution.
The first guess for each parameter was chosen from the output of SExtractor, except for the background that we fitted as a free parameter and then double checked reasonableness. In this way, we performed a first analysis leaving all the parameters free to vary ($x_0,y_0$, $K_{\rm tot}$, $R_{\rm e}$, $n$, $b/a$, and $PA$).
88.6 per cent of the early-type galaxies have $n\ge 3$ while 67.2 per cent of the late-type galaxies have $n\le 1.5$, as shown in the Figure~\ref{fig:nsers}; thus their morphology is well recovered. The mean values are $<n_{\rm early}>$=6.09 and $<n_{\rm late}>$=1.04 with standard deviations of $< \sigma_{\rm early}> $=3.9 and $< \sigma_{\rm late}>$=1.19. We were not able to decompose 17 per cent of the late-type galaxies because of their faintness: an image rebinning would increase the statistics. Galaxies with very high ellipticity, close to edge-on, have the highest errors and show a wider spread of recovered parameters, such as Obj. ID24, ID30, ID36, and ID43. In this kind of analysis, the errors in the PSF full width at half maximum (FWHM) can affect the fit and play a key role: to quantify the error coming from the assumption of only 30 SR bins, we performed the photometric analysis using the test PSFs, built with several realizations of the wavefronts within the same bin, as explained in Sections~\ref{sec:PSF}. The differences in recovering the S\'ersic parameter are always less than $3$ per cent if compared to the ones recovered using the PSF from our sample, and therefore we assumed as negligible. However, small objects with sizes comparable to the FWHM can lead to extreme values of the S\`ersic index \citep{Caon1993,Davari2016,Mendez-Abreu2016}. Their fits may truly represent the best possible S\'ersic profile, but the inferred structural parameters can be in some cases not astrophysically meaningful. This could be assessed on an object-by-object basis and through refined hand-tuned fitting, which is beyond the aim of our work. In fact since we are interested in obtaining a reliable estimate of the galaxy structure, high values of $n$ might impact considerably on other parameters. For this reason, we performed a second detailed analysis fitting a de Vaucouleurs profile (fixing $n=4$) to the objects that the first analysis has tagged as early types ($n \ge3$), and an exponential profile ($n=1$) for those where $n \le 1.5$, \ie\ tagged as late types. This is a relatively common approach used in a number of works to ensure a more robust fitting process given the reduction of the number of free parameters \citep{Trujillo2006,Lange2016,Margalef2016} and provides a good description of the average profile of early-type galaxies both nearby and distant \citep{vanderWel2008}. In our case, the differences between the parameters obtained in the two analysis are between 10 and 30 per cent, depending on the S/N of the images, and in agreement with the findings of the studies performed using real data \citep{Hopkins2009,Szomoru2010,Saracco2014}.
  
\begin{figure*}
  \begin{center}
    \includegraphics[width=0.92\hsize,angle=0]{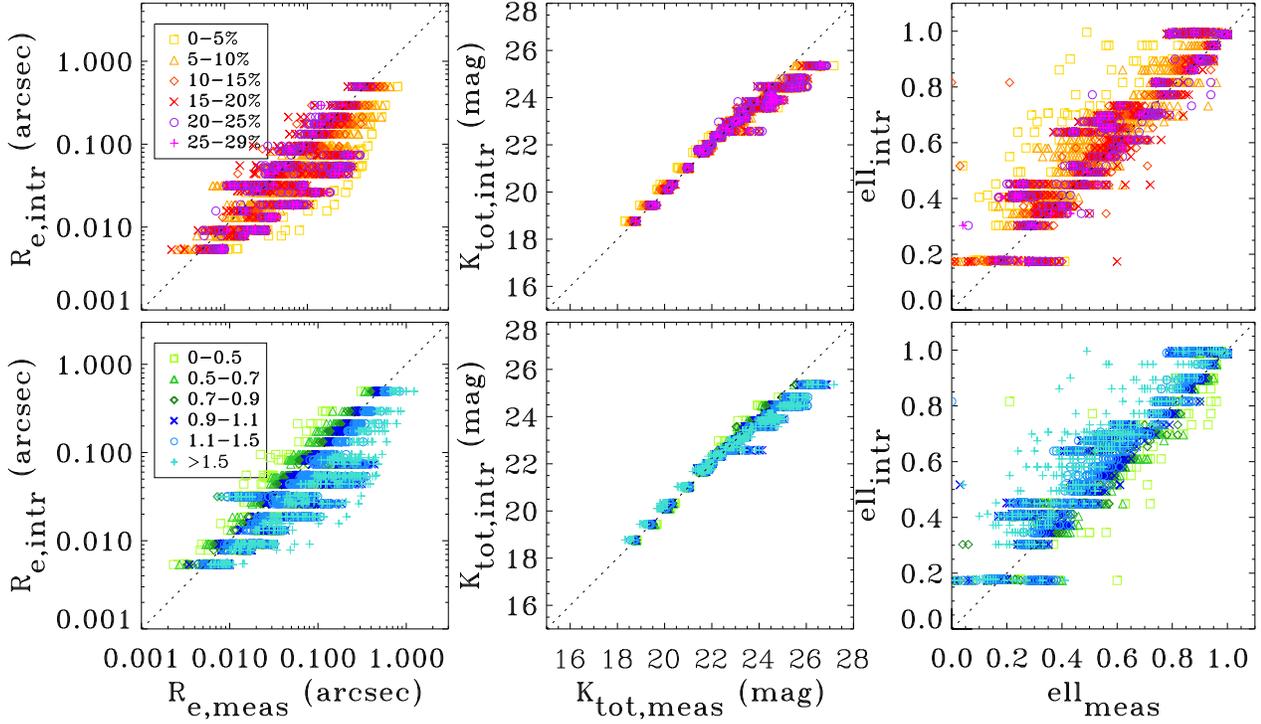}
\caption{\small{Relation between the recovered and intrinsic photometric parameters for the early-type galaxies of the sample: $R_{\rm e}$ (left-hand panel), $K_{\rm tot}$ (central panel), and $\epsilon$ (right-hand panel).
The top row represents the distribution of the SR used for convolving the image, while the bottom row considers the ratio between the mean SR of the 3 stars used to build the PSF in the field and the SR used to simulate the galaxy. The dashed line represents the bisector, \ie\ the locus where we expected to find the points. }}
\label{fig:galfit}
  \end{center}
\end{figure*}

\begin{figure*}
  \begin{center}
    \includegraphics[width=0.92\hsize,angle=0]{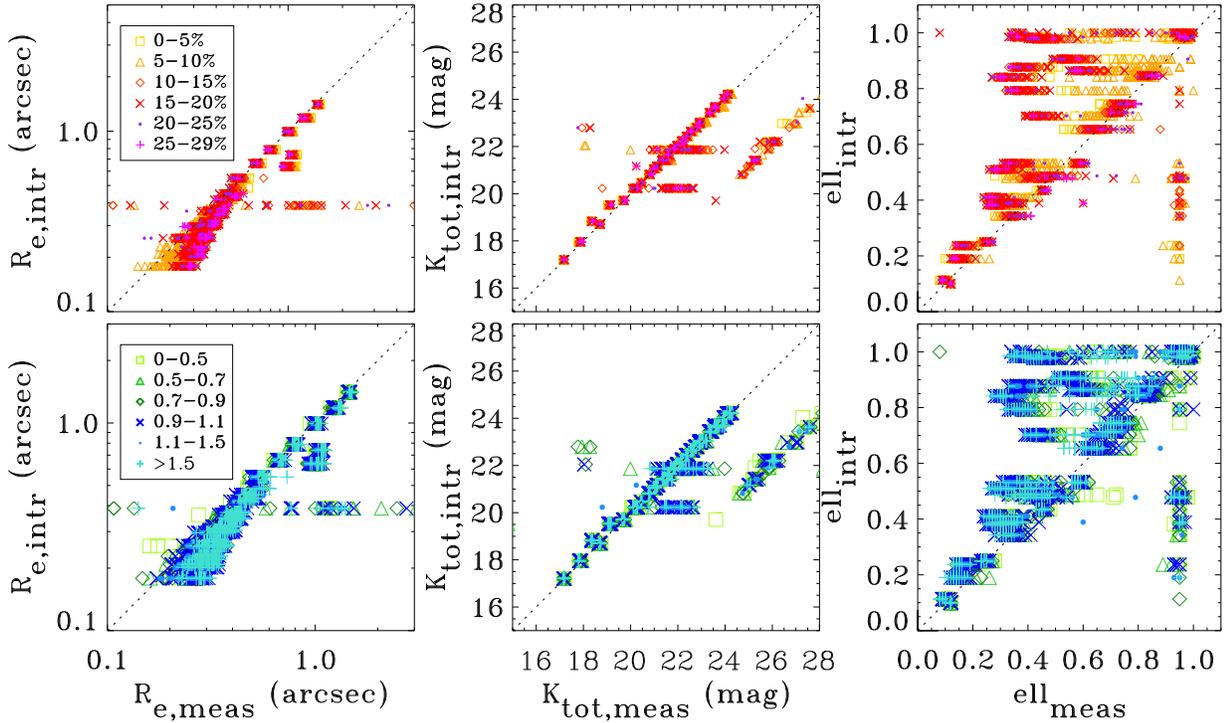}
\caption{\small{As for Figure~\ref{fig:galfit} but referring to the late-type galaxies of the sample.}}
\label{fig:galfit2}
  \end{center}
\end{figure*}

The structural parameters we obtained ($R_{\rm e,meas}$, K$_{tot,meas}$, and $ell_{\rm meas}$) were compared with the intrinsic ones ($R_{\rm e,intr}$, K$_{tot,intr}$, and $ell_{\rm intr}$), as shown in Figure~\ref{fig:galfit} and Figure~\ref{fig:galfit2}. The color code highlights the possible dependances of the results from the SR parameters: the top panel shows the SR of the PSF used to convolve the galaxy as a parameter of investigation, while the bottom panel considers the ratio between the averaged-SR of the three stars used to build the PSF in the field and the SR used to simulate the galaxy ($SR^*/SR_{\rm sim}$). The latter seems to play a key role in recovering the proper shape of the objects: the more similar the PSFs, the better the photometric analysis. 
The use of a large variety of PSFs (they are different for all the 100 fields) determines the wide spread of the values recovered. Also in this case, galaxies with very high ellipticity show the highest errors. We obviously expected to find the points along the bisector, as happens when $SR^*/SR_{\rm sim} \approx 1$.

Instead, when the PSF we used in photometric analysis has a narrower core than the one used to build the image of the simulated galaxy (thus $SR^*$ is bigger),  $SR^*/SR_{\rm sim}$  is greater than 1. In this case usually, GALFIT enlarges the effective radius of the galaxy, leading to an overestimation of this parameter and at the same time an underestimation of the ellipticity, an effect that is also described in \citet{Peng2010} and that highlights the importance of a good PSF recovery. 
 To try to overcome these systematics, one method could be to build several PSFs for each observation (or sector in our case) changing the combination of the 3 stars available in the field to check whether the results strongly change. This yealds the chance to analyse each object using PSFs that could have SRs similar to the one of galaxy. The average of the obtained parameters (or a very promising $\chi^2$ value of the fit) would represent a more robust result.
  However, although an optimization of the analysis is possible (but beyond the aim of this study), our two-dimensional decomposition of the early-type galaxies is performed successfully in all cases, showing low $\chi^2$ values. In contrast, in some cases the analysis of late-type galaxies is difficult: the low S/N of small and faint objects lead to wrong results, as is clearly visible in the outlayers that are aligned parallel to the bisector in the central panels of Figure~\ref{fig:galfit2}. These sources have been retrieved with $ell=0$, \ie\ totally flat systems, so they can be tagged as spurious decomposition. In those cases, performing a 2$\times$2 rebinning would certainly help to improve the analysis. In this context it is worth recalling that our work is pushing the present-day observational limits, which at the moment are given by HST observations of high-redshift galaxies with mass $\log M/M_{\odot}\gtrsim 10$ \citep{Brammer2012,vanderWel2012,vanderWel2014}.

\begin{table}
\caption{\small{Averaged photometric parameters of the early- ($n$=4) and late-type galaxies ($n$=1) derived from the photometric decomposition of the mock images of all the sectors.}}
\label{tab:results}
\begin{scriptsize}
\begin{center}
\begin{tabular}{crrr} 
\hline
\multicolumn{1}{c}{Morphology} &
\multicolumn{1}{c}{$\left < \frac{R_{\rm e,intr}}{R_{\rm e,meas}} \right >$} &
\multicolumn{1}{c}{ $\left < \frac{\rm{K}_{\rm tot,intr}}{\rm{K}_{\rm tot,meas}} \right >$} &
\multicolumn{1}{c}{$\left < \frac{\epsilon_{\rm intr}}{\epsilon_{\rm meas}} \right >$} \\
\hline
early type &  1.467 $\pm$ 1.2 & 1.001 $\pm$ 0.016 & 1.145 $\pm$ 0.539      \\
late type  &  1.871 $\pm$ 1.2 & 1.010 $\pm$ 0.057 & 1.904 $\pm$ 0.537      \\
\hline               
\end{tabular}
\end{center}
\begin{minipage}{8.5cm}
  NOTES --
 Averaged value of the ratio between the intrinsic and recovered effective radius (Col. 1) for all the sample objects, $K$-band total magnitude (Col. 2), and ellipticity (Col. 3).
1-$\sigma$ error is given.
\end{minipage}
\end{scriptsize}
\end{table}

\section{Summary and Conclusions}
\label{sec:concl}
We have performed end-to-end simulations building 100 mock extragalactic fields (in a grid of 10$\times$10 sectors for a total FoV of 500$\times$500 arcsec$^2$) assuming they were collected by an ELT that benefits from GMCAO, aiming at investigating the performance of such a system that uses only NGSs in a wide technical FoV.
The input parameters of the high-redshift galaxies were selected following the results of \citet{vanderWel2014}: we chose 5 masses ($\log {M/M_{\odot}}=9$,~$9.3$,~$9.7$,~$10$,~$10.3$) and 6 redshifts ($z= 0.25$, 0.75, 1.25, 1.75, 2.25, 2.75), for a total of 30 early- and 30 late-types galaxies in each sector. 
The $K$-band surface brightness of the galaxies in all fields were convolved with a set of PSFs resulting from the residual correction wavefronts given as an output of the tomographic simulation tool. Such analysis was done pointing at the CDFS, a well-known star-poor region used to carry out several important surveys, and selecting up to six stars. Our main conclusions can be summarised as follows.
\begin{itemize}
\item   The SRs achieved by the GMCAO-assisted system range from 0.01 to 0.29. The distribution over the entire field is discontinuouos in some cases, due mainly to the fact that the asterisms used to correct contiguous fields can be dramatically different. The details of the outcome of this work depend upon the turbulence profile assumed: the atmospheric model we considered is composed by 40 turbulent screens, but the use of a less pessimistic profile (like the 35-layers official one) would certainly lead us to achieve a fully optimized performance. It could be interesting to investigate whether some a-priori knowledge of the turbulence profile would improve the outcome of this work. Such information would open a wide range of possibilities, ranging from purely instrumental consequences, \ie\ a more specific optimization of the GMCAO loop, to further improvements in the analysys strategy, \ie\ the use of a-priori knowledge of the spatial frequencies of the PSFs where most of the anisoplanatism is expected. In this perspective, the comparison of the results discussed here with other competing techniques should be accomplished using uniform, or at least similar, turbulence profiles.
\item We were able to detect 99.7 and 89.4 per cent of early- and late-type galaxies of our sample, respectively, using SExtractor. To increase the S/N of the images with the late-type objects, we performed a $2\times2$ rebinning, worsening the spatial resolution but increasing up to 95.3 per cent the detection completness. Therefore GMCAO would be able to recover (low-mass) galaxies up to high redshifts even if the image has a low SR. This is very important especially because only few studies of small elliptical systems ($\log M/M_{\odot}<10$) have been carried out. 
\item  We were able to recover the morphology of the sample galaxies well using GALFIT. A comparison between the photometric parameters with the intrinsic ones reveals that the total magnitudes are in agreement in the majority of cases, while the structural parameters depend strongly on how similar the PSF used for the decomposition is to the one used to build the image. The statistics of this comparison is listed in Table~\ref{tab:results}.
\item Since the size of distant galaxies is very small (a fraction of arcsec), the PSF affects the shape of the objects strongly. This points to the necessity for a certain homogeneity of the correction of the field, associated with a good PSF reconstruction. The impact of a wrong PSF reconstruction can be severe on the measurement of the structural parameters of distant galaxies, even if a better optimization of the 2-dimensional fitting analysis might reduce it. When the modelled PSF is similar to the one used to build the image, the outcome is a rather accurate estimation of the galaxy’s astrophysical nature. Therefore, qualitatively speaking, how randomly the PSF quality changes over the FoV is more relevant than the actual SR variance. All these effects are taken into account implicitly in the simulations we carried out: in fact, detailed information on the PSF used to generate the synthetic galaxies images is not considered at all, and hence the simulations do not use any a-priori assumption. We highlight that LGSs-based MCAO systems are not immune from such difficulties, which also scale with wavelengths: SR disuniformities of a factor of 2 within a FoV similar to one sector considered in our simulations can be found also in LGSs systems \citep{Neichel2014}. However, when the AO correction is pushed toward the bluer wavelengths \citep{Hibon2016}, the feasibility of significant sky coverage with solely NGSs, together with the PSF uniformity, becomes an issue in favour of the use of LGSs.
\end{itemize}
A GMCAO-assisted telescope can carry out photometric surveys successfully, recovering the morphology and the photometry of the sample galaxies adequately. This could represent an impact science case for the new instruments on upcoming telescopes. We highlight that these results were obtained using the CDFS as a case-study, with the strong limitations that this implies. However, even within these limitations and with the choice of a pessimistic $C_n^2$ profile, which plays a crucial role on the system performance estimation, the GMCAO technique is a reliable approach to assist ELTs observations, overcoming the problems induced by LGSs and at the same time increasing the sky coverage because of the wide technical FoV.

\section{Acknowledgements}
We are especially grateful to the Referee R. Gilmozzi for the valuable suggestions that improved this manuscript 
EP thanks C. Arcidiacono, R. Falomo, L. Greggio, and L. Schreiber for the useful discussion while this paper was in progress.
This work has been supported by the Italian Ministry for Education University and Research (MIUR) under the grant {\it T-REX2 Premiale ELT 2013}.
\bibliographystyle{mn2e1}
\bibliography{708main}{}

\label{lastpage}

\end{document}